\documentclass[twocolumn,prapplied,aps,amsmath,showpacs,floatfix,superscriptaddress]{revtex4-2}

\usepackage{graphicx}
\usepackage{dcolumn}
\usepackage{mathtools}
\usepackage{color}
\usepackage{braket}
\usepackage{hyperref}
\graphicspath{{figures/}}
\usepackage{soul}

\renewcommand{\S}{S_{11}}
\newcommand{\Qe}{Q_{ext}}
\newcommand{\Qi}{Q_{int}}
\newcommand{\fres}{f_{0}}
\newcommand{\Zc}{Z_{char}}
\newcommand{\vn}{\sigma_v}

\newcommand{\facc}{f_{0}^\mathrm{acc}}
\newcommand{\fdepl}{f_{0}^\mathrm{depl}}

\newcommand{\SNR}{\mathrm{SNR}}

\newcommand{\sqHz}{\sqrt{\mathrm{Hz}}}

\newcommand{\qutech}{QuTech and Kavli Institute of Nanoscience, Delft University of Technology, 2600 GA Delft, The Netherlands}
\newcommand{\msdelft}{Microsoft Quantum Lab Delft, Delft University of Technology, 2600 GA Delft, The Netherlands}
\newcommand{\TUe}{Department of Applied Physics, Eindhoven University of Technology, 5600 MB Eindhoven, The Netherlands}
\newcommand{\CPH}{Center for Quantum Devices, Niels Bohr Institute, University of Copenhagen \& Microsoft Quantum Materials Lab Copenhagen, Lyngby, Denmark}

\begin{document}
\title{Radio-frequency C-V measurements with sub-attofarad sensitivity}

\author{Filip~K.~Malinowski}
\email{f.k.malinowski@tudelft.nl}
\affiliation{\qutech}

\author{Lin~Han}
\affiliation{\qutech}

\author{Damaz~de~Jong}
\affiliation{\qutech}

\author{Ji-Yin~Wang}
\affiliation{\qutech}

\author{Christian~G.~Prosko}
\affiliation{\qutech}

\author{Ghada~Badawy}
\affiliation{\TUe}

\author{Sasa~Gazibegovic}
\affiliation{\TUe}

\author{Yu~Liu}
\affiliation{\CPH}

\author{Peter~Krogstrup}
\affiliation{\CPH}

\author{Erik~P.A.M.~Bakkers}
\affiliation{\TUe}

\author{Leo~P.~Kouwenhoven}
\affiliation{\qutech}
\affiliation{\msdelft}

\author{Jonne~V.~Koski}
\affiliation{\msdelft}

\date{\today}

\begin{abstract}
	We demonstrate the use of radio-frequency (rf) resonators to measure the capacitance of nano-scale semiconducting devices in field-effect transistor configurations.
	The rf resonator is attached to the gate or the lead of the device. Consequently, tuning the carrier density in the conducting channel of the device affects the resonance frequency, quantitatively reflecting its capacitance.
	We test the measurement method on InSb and InAs nanowires at dilution-refrigerator temperatures.
	The measured capacitances are consistent with those inferred from the periodicity of the Coulomb blockade of quantum dots realized in the same devices.
	In an implementation of the resonator using an off-chip superconducting spiral inductor we find sensitivity values reaching down to 75~zF/$\sqHz$ at 1~kHz measurement bandwidth, and noise down to 0.45~aF at 1~Hz bandwidth.
	We estimate the sensitivity of the method for a number of other implementations. In particular we predict typical sensitivity of about 40~zF/$\sqHz$ at room temperature with a resonator comprised of off-the-shelf components.
	Of several proposed applications, we demonstrate two: the capacitance measurement of several identical 80~nm-wide gates with a single resonator, and the field-effect mobility measurement of an individual nanowire with the gate capacitance measured in-situ.
\end{abstract}

\maketitle

\section{Introduction}

Radio-frequency (rf) resonators are broadly used for readout of solid-state qubits, whether the quantum information is encoded as an excitation of a superconducting circuit~\cite{krantz2019}, a quantum dot's charge, or an electronic spin~\cite{chatterjee2021}. This follows from mappings of the equivalent resistance, capacitance or inductance of the quantum system to the transmission or reflection coefficient of the macroscopic rf resonator. Furthermore, resonance frequencies between tens of megahertz and tens of gigahertz enable the use of the near-quantum-limited cryogenic amplifiers, with noise temperatures below 1~K~\cite{macklin2015,schupp2020}. Seminal work of Schoelkopf~et~al.\cite{schoelkopf1998} demonstrated an orders-of-magnitude improvement in electrometer sensitivity, once the low-frequency measurement of the single-electron transistor is substituted by embedding it in an rf resonator.

Concurrently, a number of methods have been developed to measure capacitance of field-effect transistors (FETs), Schottky junctions and various other semiconducting devices\cite{gunawan2008,garnett2009,tseng2010,ji2013}. A common feature of these measurement schemes is the use of relatively low excitation frequencies, up to tens of kilohertz, raising a question of whether a scheme similar to Ref.~\onlinecite{schoelkopf1998} could be applied to capacitance measurements. Such an approach would increase the measurement bandwidth and reduce the influence of the ubiquitous $1/f$ noise by several orders of magnitude. Instead, the measurement would become limited by much smaller Johnson–Nyquist noise, amplifier noise and noise intrinsic to the device.

In this work we demonstrate the use of rf resonators to measure the bulk capacitance of micro- and nanometer-scale devices. The rf resonator is attached to a source or gate of the semiconducting nanowire device in FET configuration, and its resonance frequency is measured as a function of gate voltages. We demonstrate that the frequency shift quantitatively matches the gate capacitance.
The sensitivity and bandwidth of our capacitance measurement scheme exceed the capability of alternative methods by more than an order of magnitude\cite{ilani2006,tseng2010,ji2013} and are consistent with the recent use of an rf resonator to measure the quantum capacitance of a quantum point contact~\cite{jarratt2020}. With such a high sensitivity, moderate integration times ($<$1~s) are sufficient to measure capacitances of devices with a footprint below 100~nm, as well as quantum contributions to a bulk capacitance of other devices with a typical scale of $\sim$1~$\mu$m \cite{ilani2006,jarratt2020}.

In section~\ref{method} we introduce the principle of the capacitance measurement with an rf resonator, and discuss its variants and limitations. Section~\ref{validation} presents a validation of the method by comparing the capacitance extracted from the resonator shift with the one extracted from the periodicity of Coulomb blockade in three devices with different gate sizes. In section~\ref{sensitivity} we characterize the sensitivity and the noise of the capacitance measurement, and estimate the performance of the method for several different temperatures, amplifiers, and realizations of rf resonators. Section~\ref{applications} presents the application of the method to measure uniformity of nominally identical gates and to measure field-effect mobility of an individual device. We also discuss several other possible applications and implementations.

\section{Method}
\label{method}

The measurement method is based on the elementary fact, that the resonance frequency of an LC circuit depends on its capacitance. Therefore, if the inductance $L$ embedded in the circuit is known, one can infer the capacitance from a measurement of the resonance frequency. To this end, the resonance frequency is extracted from the frequency-dependent reflection or transmission through the resonator. In this section we describe how to apply this principle to measure the capacitance of mesosopic devices. We start by describing the devices used for the validation of the method. Next we outline the experimental procedure. We conclude the section with a discussion of a number of critical factors that need to be taken into account in order for the method to yield valid results.

\begin{figure}[tb]
	\includegraphics[scale=1]{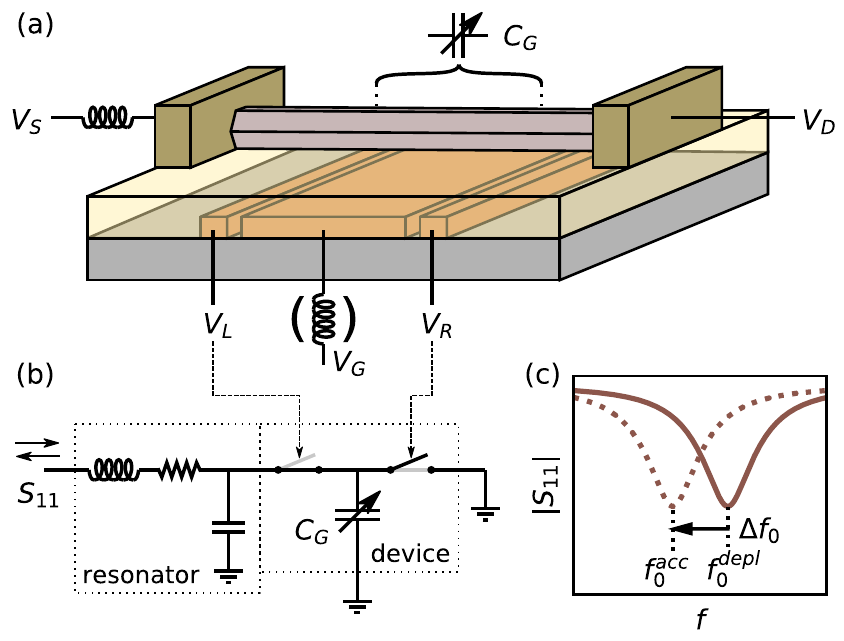}
	\caption{(a) Schematic of a bottom-gated nanowire device in a FET configuration. A source-drain voltage $V_{SD}$ can be applied to measure the conductance of the device. Gate voltages $V_{L/G/R}$ control the electron density in the semiconductor. $C_G$ symbolizes the variable capacitance between the gate and the segment of the nanowire tuned by the gate voltage $V_G$. Coil symbols indicate contacts that can be connected to an rf resonator to perform capacitance measurements.
	(b) Schematic of the circuit for the capacitance measurement with the resonator attached to the device source contact. A lossy resonator is represented as an RLC circuit. The device is represented as two switches (tuned by gate voltages $V_{L/R}$) and a variable capacitor. For a capacitance measurement the gate voltage VL is set to accumulate charge carriers, while Vr is set to deplete them.
	(c) Illustration of the resonance frequency shift, measured in reflectometry, resulting from the accumulation of electrons in the nanowire. In the measured nanometer-scale devices the frequency shift is typically much smaller than the resonance linewidth.
	}
	\label{fig_device}
\end{figure}

The nanostructures used for validation and benchmarking of the method are InSb and InAs nanowire single-electron transistors, depicted schematically in Fig.~\ref{fig_device}(a). The nanowires were grown by metalorganic vapour-phase epitaxy (InSb)\cite{badawy2019} and molecular beam epitaxy (InAs)\cite{krogstrup2015}, and deposited on a highly resistive Si substrate. The devices are bottom-gated in the case of InSb, and top-gated in the case of InAs. The Ti/Au gates for tuning the electron density in the bulk of the nanowire are defined by e-beam lithography and are isolated from the wire by $\sim$15~nm of ALD AlO$_\mathrm{x}$. Source and drain contacts are made of Ti/Au and deposited with a direct contact to the nanowire. In these devices we aim to characterize the capacitance $C_G$ between the central gate and the nanowire, as a function of the applied voltage $V_G$. For that purpose we wire-bond the source or the central gate of each measured device to superconducting NbTiN spiral inductor resonators\cite{hornibrook2014} with inductances between 420 and 730~nH. Together with the parasitic and their self-capacitance (typically 0.3-0.35~pF) these inductors form LC circuits with resonance frequencies between 300 and 500~MHz. The side gate voltages, $V_{L/R}$, locally tune the electrostatic potential in order to connect the bulk of the wire to the contacts, isolate it, or to form tunnel barriers.

Fig.~\ref{fig_device}(b) depicts a schematic of the circuit for the capacitance measurement with the rf resonator connected to the source of the device. Voltage $V_L$ is set to provide a good galvanic connection\footnote{Good galvanic connection in this context means that the contact resistance $R$ must be such that $2 \pi/R C_G \gg \fres$. For $\fres = 500$~MHz and $C_G = 1$~fF this implies $R \ll 10$~M$\Omega$.} between the contact and the tuneable bulk of the nanowire. Voltage $V_R$ is such that the nanowire is locally fully depleted to disconnect the drain contact which would otherwise act as an rf ground. Next, the reflection from the resonator is measured to extract the resonance frequency as a function of the gate voltage $V_G$. Fig.~\ref{fig_device}(c) illustrates schematically that in the reflection measurement the resonance frequency is reduced when the nanowire is accumulated (dashed line), relative to when the charge carriers are depleted (solid line).

Using the known, designed value of inductance $L$, and having measured the resonance frequency with the carriers depleted and accumulated $\fres^\mathrm{depl/acc}$, the gate capacitance $C_G$ is obtained from a set of two equations:
\begin{equation}
	\fdepl = \frac{1}{2\pi \sqrt{LC}}; \ 
	\facc = \frac{1}{2\pi \sqrt{L(C+C_G)}},
\end{equation}
where $C$ is the capacitance of the resonator, including the self-capacitance of the inductor and parasitic capacitance of the bondwires and the source contact. Equivalently
\begin{align}
	C_G & = \frac{1}{(2 \pi \fdepl)^2 L} - \frac{1}{(2 \pi \facc)^2 L} \\
		& \approx -\frac{\Delta \fres}{2\pi^2 (\fdepl)^3 L}, \notag
\end{align}
where the last approximation assumes a small resonator shift $\Delta \fres \ll \fres$.

With the resonator attached to the device gate the capacitance measurement is performed similarly. In this case, either one or both of the barrier gate voltages $V_{L/R}$ must be set to accumulate electrons to provide a low-impedance shunt to rf ground. The two approaches: with the resonator connected to the device source or gate, measure slightly different quantities, and are subject to different constraints.

First, the rf resonance frequency is affected by capacitance between the attached contact (source or gate) and any rf ground. This includes e.g. other gates or a conducting substrate. If gating of the active part of the device also influences other relevant parts of the device (e.g. the substrate), the capacitance method becomes unreliable. In the validation experiment we ensure that is not the case by using a highly resistive Si substrate.

Second, the channel tuned by $V_G$ may be coupled capacitatively to multiple gates. Thereby, resonators attached to the device lead or gate will measure different capacitance. If the resonator connected to the tuned gate, it will only detect capacitance changes between the channel and the tuned gate. Meanwhile, a measurement with the resonator connected to the source lead detects capacitance changes between the carriers accumulated in the channel and \emph{all} gates. Depending on the purpose of the measurement either one or both of these implementations may be desirable. In our devices the capacitance between the channel and the gate dominates over capacitances between the channel and other elements of the device.

Third, attaching the resonator to the device source enables using it for measuring the capacitance of multiple connected devices, since the DC gate voltages can be applied separately, and used to select between the devices. However, if the device yield is low this may be undesirable, as a single faulty device may introduce an rf short to ground and render the resonator unusable.

Finally, in case the resonator is attached to the device source it can be repurposed for rf conductance measurements \cite{schoelkopf1998,razmadze2019,dejong2021}. When the source-drain resistance is finite, it introduces dissipation in the resonant circuit, mapping the channel conductance to the internal quality factor of the resonator.

In the validation experiment (Sec.~\ref{validation}) we measure the capacitance of three InSb devices: for two of them we attach the resonator to the source contact, and for the last one -- to the gate. In a noise characterization experiment (Sec.~\ref{sensitivity}) we measure the resonator attached to the source contact of InAs device.

\section{Validation}
\label{validation}

\begin{figure}[bt]
	\includegraphics[scale=1]{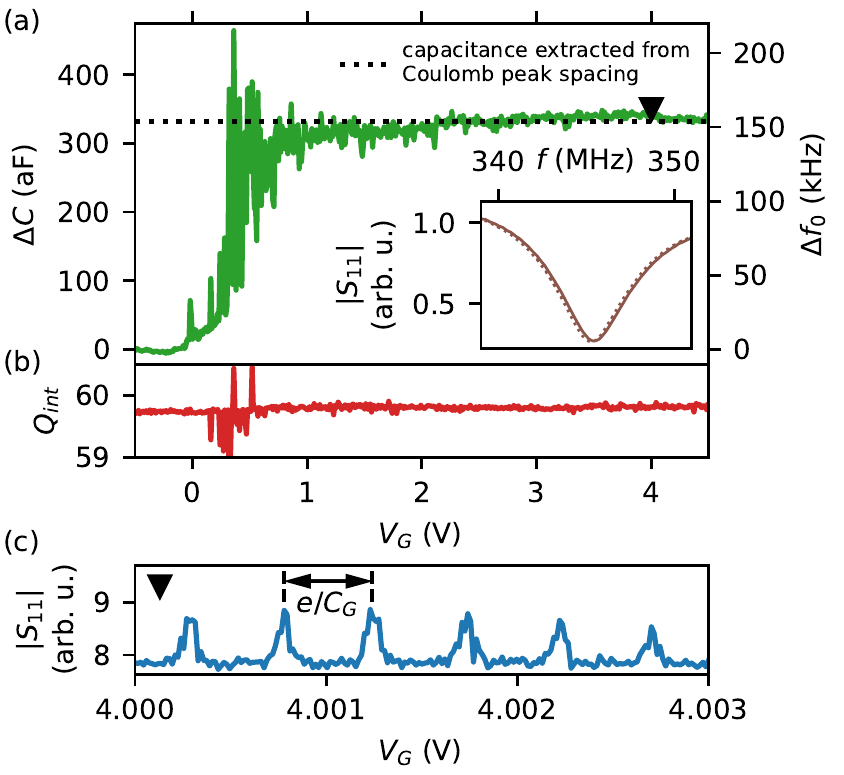}
	\caption{(a) Frequency shift, $\Delta f$, and a corresponding capacitance change, $\Delta C$, as a function of the plunger gate voltage $V_G$ for 2-$\mu$m long InSb device. Dotted line indicates the gate capacitance extracted from Coulomb blockade measurement at $V_G=4$~V, indicated with a triangular marker. Inset: $|\S|$ versus frequency at $V_G = -0.5$~V (solid line) and $4$~V (dotted).
	(b) Internal quality factor $\Qi$ of the resonator.
	(c) Rf Coulomb blockade measurement at zero DC bias. Coulomb peak spacing provides an independent measure of the gate capacitance.}
	\label{fig_comparison}
\end{figure}

To verify that the rf capacitance measurement provides quantitatively accurate results, we extract capacitance from the resonator shift and from the periodicity of the Coulomb blockade for InSb devices with 80, 500 and 2000~nm gate widths. An example of such a comparison is shown in Fig.~\ref{fig_comparison}, for a device with 2~$\mu$m-wide bottom gate. 

First, we pinch off a section of the nanowire with a gate voltage $V_R$ and accumulate with $V_L$, thereby connecting the nanowire bulk to the source contact and disconnecting it from the drain. Next, we measure the reflection $\S$ from the rf resonator as a function of drive frequency $f$ and gate voltage $V_G$. For each $V_G$, we fit the resonator response (see Appendices~\ref{app_sensitivity}, \ref{app_asymmetric}, \ref{app_fit}). The extracted resonance frequency change, $\Delta \fres$, capacitance change, $\Delta C$, and resonator internal quality factor, $\Qi$, are plotted in Fig.~\ref{fig_comparison}(a,b).

The resonance frequency is nearly constant below about $V_G\approx -0.2$~V. The resonator shift gradually increases when increasing $V_G$ from $-0.2$ to $1$~V, and saturates above $V_G \approx 1$~V. The total shift of about 150~kHz corresponds to an increase in capacitance of $C_G = 334.8 \pm 1.0$~aF, which we identify as a gate-wire capacitance. The range over which the capacitance gradually increases is consistent with the range over which the gate voltage $V_G$ tunes the channel from fully closed to fully open in a lock-in conductance measurement (c.f. Sec.\ref{mobility} and Fig.~\ref{fig_mobility}). We attribute the peaks in the intermediate range of $V_G$ to the quantum capacitance of unintentional quantum dots formed in the disordered potential in the wire. Panel (b) demonstrates the change of the resonator internal quality factor $\Qi$ as a function of the gate voltage. The inset of Fig.~\ref{fig_comparison}(a) illustrates a small change of the resonance frequency compared to the resonator linewidth. 

To independently measure the gate capacitance with a different method, we form a quantum dot by setting the gate voltages $V_{L,R}$ to a tunneling regime ($\sim$150~mV), and $V_G\approx 4$~V. Coulomb peaks versus $V_G$ are measured by means of rf-conductance\cite{schoelkopf1998, razmadze2019, dejong2021}, using the same resonator that was used for the direct capacitance measurement [Fig.~\ref{fig_comparison}(c)]. The spacing between the Coulomb peaks $\Delta V_G$ corresponds to the gate voltage change required to add a single electron to a quantum dot\cite{ihn2010}, and is related to the gate capacitance via $\Delta V_G = e/C_G$. This measurement yields capacitance $C_G = 331.7 \pm 7.8$~aF, in good agreement with the value extracted from the resonator shift.

The same measurement is repeated for two other InSb nanowire devices, with gate widths of 80 and 500~nm. Table~\ref{tab_comparison} lists the capacitances extracted by both methods for all three devices, and supporting data is presented in Appendix~\ref{app_bonus}. In all cases we find that the two methods are in good agreement.

\begin{table}
\caption{Comparison of gate capacitance extracted from Coulomb blockade and resonator shift.}
\label{tab_comparison}
\begin{tabular}{c||c|c|c}
 & Resonator & Capacitance (aF) & Capacitance (aF) \\
Gate size & placement & Resonator shift & Coulomb blockade \\ 
\hline  \hline 
80 nm & Source & 31.7$\pm$0.6 & 27.5$\pm$1.7 \\ 
\hline 
500 nm & Gate & 115.1$\pm$0.2 & 114.4$\pm$3.7 \\ 
\hline 
2000 nm & Source & 334.8$\pm$1.0 & 331.7$\pm$7.8 \\ 
\end{tabular} 
\end{table}

\section{Sensitivity}
\label{sensitivity}

\begin{figure}[tb]
	\includegraphics[scale=1]{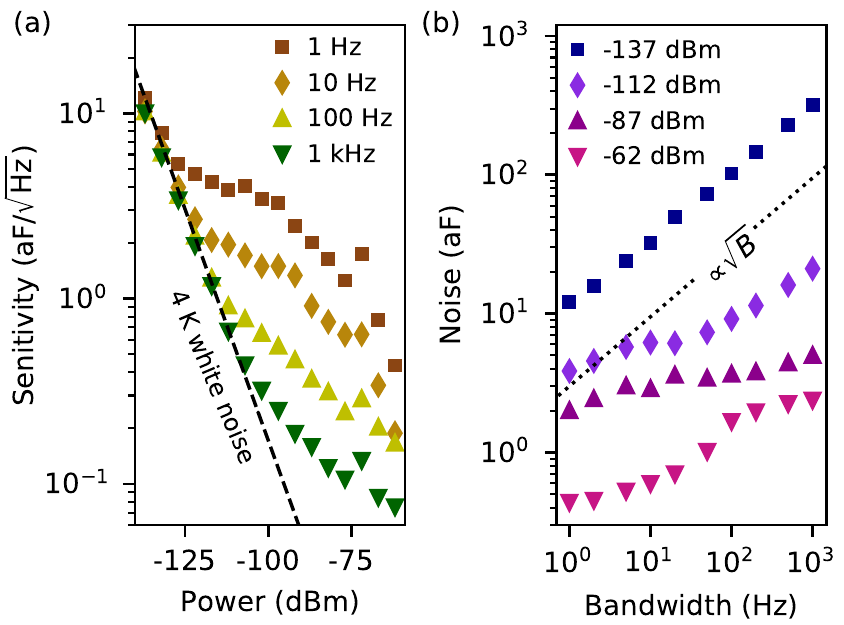}
	\caption{(a) The measured sensitivity of the capacitance measurement with a resonator. The dashed line indicates the sensitivity limit due to the amplifier noise ($T_\mathrm{noise} \approx 4$~K).
	(b) The standard deviation of the gate capacitance measurement vs bandwidth. The noise saturates at about 0.4-1~aF due to drift over a 4.5-minute-long measurement. The dotted line indicates the scaling of the noise $\propto \sqrt{B}$, expected for white noise such as Johnson-Nyquist noise.}
	\label{fig_sensitivity}
\end{figure}

Next, we investigate the sensitivity of the capacitance measurement. To maximize the sensitivity it is beneficial to perform the measurement only near the resonance frequency. With an analytical model of the resonator and assuming fixed quality factors, the amplitude and phase measurement at fixed frequency allows us to determine the frequency shift and, therefore, the added capacitance (Appendix~\ref{app_fit}, \ref{app_fixed_freq}).

We characterize the sensitivity of the capacitance measurement on an InAs nanowire device with a 340~nm wide gate wrapped around the nanowire. The inductor ($L = 730$~nH) is attached to the source lead of the device. The resonator has a resonant frequency $\fres = 312.81$~MHz, an internal quality factor of $\Qi = 488$, and an external quality factor of $\Qe = 134$.

We acquire time traces of the signal reflected from the resonator at a fixed frequency $f=313.21$~MHz, first with the gate voltage $V_G$ set to fully open, and then to fully closed. This procedure is repeated for various measurement bandwidths $B$ and drive powers, while maintaining a fixed total 4.5~min length of the time traces. The fixed duration of several minutes is chosen in order to expose the setup to detrimental influence of the device drift and low-frequency noise, and to obtain fair values of a noise level applicable to real use case. We define the SNR for distinguishing the opened and closed cases as
\begin{equation}
	\SNR = \left\lvert \frac{\S^\mathrm{open} - \S^\mathrm{closed}}{(\sigma_{11}^\mathrm{open} + \sigma_{11}^\mathrm{closed})/2} \right\rvert.
\end{equation}
where $\S^\mathrm{open/closed}$ and $\sigma_{11}^\mathrm{open/closed}$ are the mean and standard deviation of the measured complex reflection coefficient for the gate being opened and closed. Based on the $\SNR$ we define the noise and the sensitivity of the capacitance measurement as $N = \Delta C / \SNR$ and $s = \Delta C / (\SNR \times \sqrt{B})$, respectively.

Fig.~\ref{fig_sensitivity} shows the measured sensitivity and noise for various measurement bandwidths and excitation powers. Panel (a) demonstrates that the measurement sensitivity improves with increased rf excitation power. The low-power sensitivity is independent from the measurement bandwidth. As power is increased, the sensitivity becomes increasingly more bandwidth-dependent with the lowest values corresponding to the highest used bandwidth of 1~kHz. In Fig.~\ref{fig_sensitivity}(a) we indicate the sensitivity expected for $\sim$4~K thermal (white) noise (Appendix~\ref{app_sensitivity}), which approximates the expected noise level for the used cryogenic amplifier (Cosmic Microwave CITLF2, nominal noise temperature: 1.6~K) mounted at the 4~K plate of the dilution refrigerator (actual temperature: 3.4~K). For 1~kHz bandwidth and below $-100$~dBm drive power, the measured sensitivity matches the 4~K noise limitation. When the drive power is increased further, the sensitivity starts exceeding that limit more significantly, indicating that other factors come into play. As the bandwidth is reduced, the sensitivity exceeds the 4~K limit at lower power -- for 1~Hz bandwidth the deviation from amplifier limit occurs already at about $-125$~dBm. Furthermore, a plateau-like region in sensitivity appears between about $-110$ and $-100$~dBm.

To gain insight into the absolute precision of the capacitance measurement we plot the absolute noise level in Fig.~\ref{fig_sensitivity}(b). At the lowest drive powers the noise scales as $\sqrt{B}$, indicating that white amplifier and thermal noise are dominating. As the power is increased, the noise appears to saturate at a few-attofarad level. At $-87$~dBm the bandwidth change from 1~kHz to 1~Hz only reduces the noise from 4 to 2~aF. At even higher power ($-62$~dBm) the noise decreases further, reaching the lowest value of 0.45~aF for 1~Hz bandwidth.

We speculate that the sensitivity plateaus for intermediate power at a few-attofarad level, as well as the plateaus in sensitivity represent fluctuations in the electrostatic environment of the nanowire. For low powers and high bandwidths, the fluctuations are not resolvable due to white noise. On the other hand, when the drive power increases further, these fluctuations are averaged out, and the noise level decreases further until it becomes limited by $1/f$ noise.

\begin{table*}
\caption{Summary of the estimated sensitivities for several realizations of the rf resonant circuits and amplifiers, assuming the performance is limited by Johnson–Nyquist noise.}
\label{tab_estimates}
\begin{tabular}{c|c|c|c|c|c|c|c|c|c}
Resonator type & Amplifier & $\fres$ & $\Zc$ & $\Qi$ & $\Qe$ & $P$ & $T_{noise}$ & $S_C$ & Source \\ 
 &  & (GHz) & ($\Omega$) &  &  & (dBm) & (K) & (zF/$\sqHz$) &  \\
\hline \hline 
Superconducting spiral inductors & 4 K HEMT & 0.35 & 1000 & 500 & 300 & $-100$ & 4 & 80 & this work\footnote{typical values from among several spiral inductor chips and nanowire devices} \\ 
\hline 
Superconducting CPW resonator & TWPA & 5 & 50 & $10^4$ & $10^3$ & $-120$ & 0.6 & 60 & Ref.~\onlinecite{macklin2015} \\ 
\hline 
Surface mount LC, cryogenic & Low-noise SQUID & 0.196 & 275 & 30 & 30 & $-80$ & 0.6 & 110 & Ref.~\onlinecite{schupp2020} \\ 
\hline 
Surface mount LC, room-temperature & 300 K HEMT & 0.26 & 650 & 30 & 150 & $-40$ & 300 & 40 & Appendix~\ref{app_bonus}\footnote{self-resonance of a 1~$\mu$H surface-mount inductor (Coilcraft, 1008CS-102X\_E\_)} \\ 
\end{tabular} 
\end{table*}

Finally, we estimate the expected sensitivity of the capacitance measurements for several implementations of rf resonators and noise temperatures (Table~\ref{tab_estimates}). The first entry represents a slight improvement that could be expected in a setup identical to ours, thanks to an increase of the resonator internal and external quality factors, but at moderate drive powers. The second entry shows typical sensitivities that could be achieved using the resonator design (i.e. 50~$\Omega$, superconducting CPW resonator) and the amplification chain widely used for readout of superconducting qubits\cite{krantz2019}. The third line indicates the possible performance which can be achieved using the self-resonance of off-the-shelf surface-mount inductors at cryogenic temperatures and state-of-the-art amplification\cite{schupp2020}. In the final entry, we estimate that similar sensitivity can be achieved at room temperature, when the drive power no longer needs to be limited to minimize the heating of the device above cryogenic temperatures. We note that in all cases the actual noise floor may be limited by intrinsic properties of the device under study, or $1/f$ noise, as is the case in our experiment.

\section{Examples of application}
\label{applications}

In this section we list several use cases for the capacitance measurement with rf resonators. We demonstrate how a single resonator can be used to measure capacitance of multiple gates (Subsec.~\ref{series}), and to supplement conductance measurements in extracting the mobility of an individual nanowire (Subsec.~\ref{mobility}). In Subsec.~\ref{compressibility} we list a few quantum-mechanical phenomena that affect the electronic compressibility, and thereby can be detected with a sufficiently sensitive capacitance measurement. Subsec.~\ref{probe} proposes an implementation of the rf resonator on the needle of a micromanipulator for rapid capacitance measurements in a probe station.

\subsection{Capacitance of multiple gates measured with a single resonator}
\label{series}

\begin{figure}[tb]
	\includegraphics[scale=1]{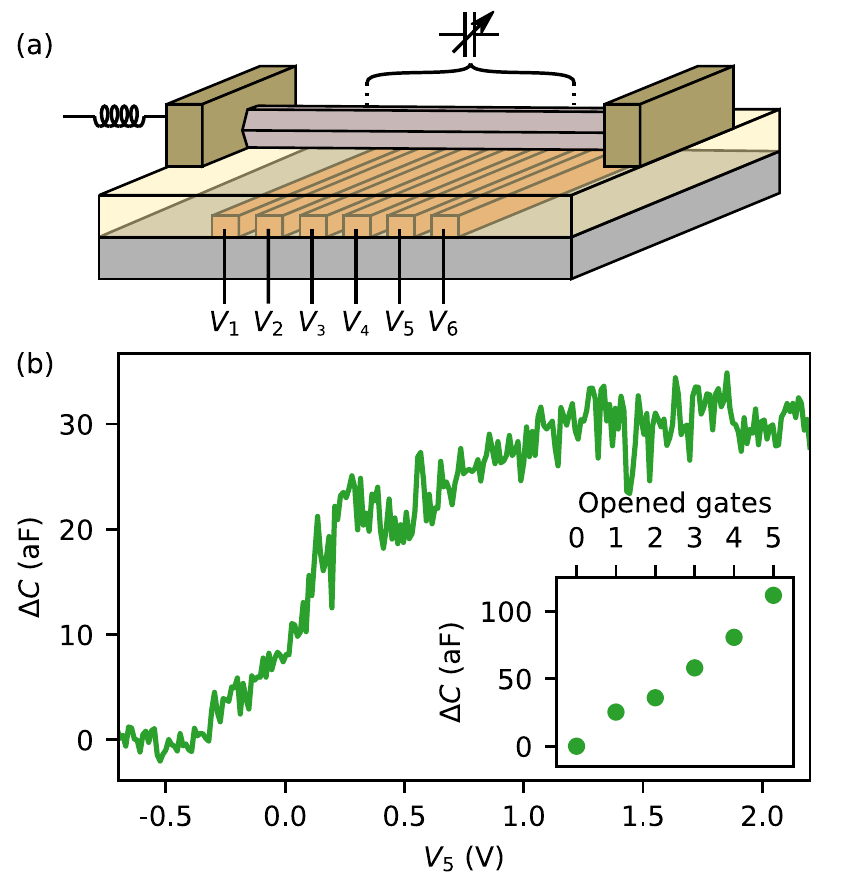}
	\caption{(a) A schematic of the InSb nanowire device with six parallel 80-nm-wide bottom gates, spaced by 60 nm. The coil symbol indicates the drain contact connected to a rf resonator.
	(b) Capacitance added to the circuit as a function of the voltage $V_5$ on one of the gates. (inset) Values of the added capacitance with 0 to 5 of the bottom gates in the open state.}
	\label{fig_series}
\end{figure}

To demonstrate that a single rf resonator attached to a lead can be used to measure multiple gates we focus on an InSb nanowire device with six 80-nm-wide parallel bottom gates [Fig.~\ref{fig_series}(a)], labeled $V_1$ through $V_6$. The rf resonator is only attached to a single lead, and yet we use it to measure the capacitance of gates 1 through 5. We start with all gates at large negative voltages. The gate voltage $V_1$ is then gradually increased, while measuring the resonance frequency of the resonator. When the measured value of the capacitance saturates we assign the corresponding value to the first gate, and proceed to sweeping the next gate voltage. The measured change of the capacitance in a sweep of a gate voltage $V_5$ is presented in Fig.~\ref{fig_series}(b). The inset of Fig.~\ref{fig_series}(b) illustrates the saturation values of capacitance with the first $N$ gates opened. In this configuration it is not possible to measure the capacitance of the sixth gate $V_6$. As soon as it accumulates carriers, the rf circuit becomes terminated with a low impedance of the drain lead and the resonance feature in a reflection measurement vanishes.

We suggest that the capability of measuring capacitance of multiple small gates with a single resonator may be applicable in development and characterization of multi-gate structures, e.g. arrays of quantum dots for spin-qubits \cite{zajac2016,veldhorst2017,chanrion2020}.

\subsection{Mobility}
\label{mobility}

\begin{figure}[tb]
	\includegraphics[scale=1]{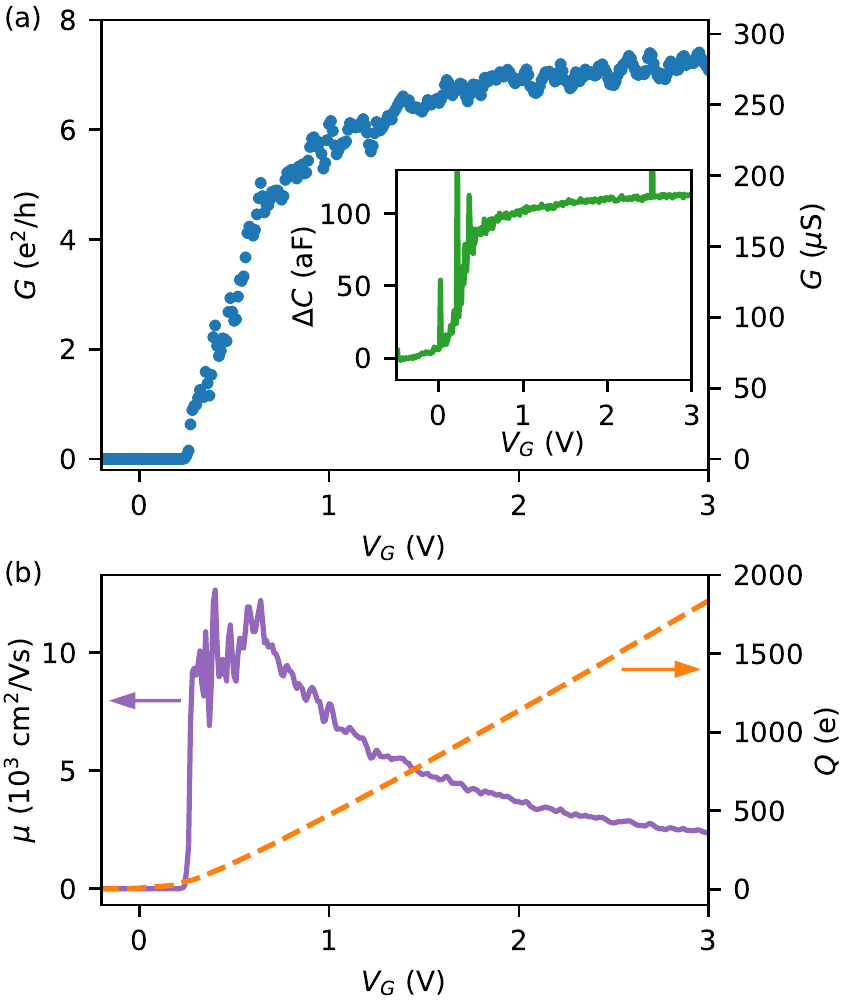}
	\caption{(a) Conductance $G$ as a function of gate voltage $V_G$ in an InSb nanowire device with 500-nm-wide bottom gate. (inset) Added capacitance $\Delta C$ as a function of gate voltage $V_G$.
	(b) Mobility of the nanowire device and charge in the conductive channel versus gate voltage $V_G$.}
	\label{fig_mobility}
\end{figure}

Next, we demonstrate the possibility of using the rf-resonator-based capacitance measurement to complement DC conductance measurements in determining the mobility of individual sub-micrometer devices. The capacitance of such devices is too small to be obtained via conventional, low-frequency C-V measurements. This is usually resolved by measuring the capacitance of multiple nominally identical devices, or by relying on finite-element simulations. These approaches may obscure the variation between individual devices or lead to systematic errors.

In our demonstration we focus on an individual InSb nanowire with a 500-nm-wide bottom gate [cf.~Fig.~\ref{fig_device}(a)]. We start by measuring the gate capacitance as described in section~\ref{method} (inset of Fig.~\ref{fig_mobility}). Afterwards, we measure the conductance across the device versus gate voltage $V_G$ with all gates opened, using a 2-terminal lock-in measurement with a 3~mVrms excitation voltage, corrected for resistances in the filtered lines of the cryostat [Fig.~\ref{fig_mobility}(a)].

We integrate numerically the gate capacitance versus the gate voltage to calculate the total charge $Q$ in the conductive channel
\begin{equation}
	Q (V_G) = e \hspace{-10pt} \int \limits_{-0.5 \mathrm{V}}^{V_G} \hspace{-10pt} \Delta C (\tilde{V}_G) \mathrm{d}\tilde{V}_G,
\end{equation}
where e is the electron charge. Finally, we calculate the mobility $\mu(V_G) = L^2 G(V_G) / Q(V_G)$, where $L=500$~nm is the nanowire length, and $G$ is the measured conductance. The calculated charge and mobility are plotted in Fig.~\ref{fig_mobility}(b). We find a peak mobility of $\mu \approx 1.2 \times 10^4$~cm$^2$/Vs. This value is somewhat lower than other measurements on nanowires grown by the same process~\cite{badawy2019}, due the use of field-effect model\cite{gul2015} to fit the data in Ref.~\onlinecite{badawy2019}, and possibly due to more involved fabrication of the devices for our experiment. The field-effect model assumes gate-independent mobility and saturation of the pinch-off curve to additional in-line (e.g. contact) resistance, yielding higher values of the extracted mobility. Using the field-effect model we extract a mobility of $\mu_{FE} = 2.3 \times 10^4$~cm$^2$/Vs (Appendix~\ref{app_bonus}).

\subsection{Electronic compressibility in mesoscopic devices}
\label{compressibility}

In multiple solid state physics phenomena, the charging of the mesoscopic system is not only affected by the geometrical capacitance of the device. Properties, like the electronic band structure and electron-electron interaction, quantified by the physical parameter of electronic compressibility, also affect the device capacitance. To date, measurements of the bulk electronic compressibility were mostly performed by means of capacitance bridges\cite{ilani2006,tomarken2019} or the electric field penetration technique\cite{eisenstein1992,smith2011,young2012}. Quantum effects of 1D and mesoscopic devices are at the very limit of what is possible to measure with these methods. With sub-attofarad noise, for the same or smaller excitation amplitudes, a number of phenomena can be further explored.

Electronic compressibility is a hugely informative quantity in topics such as Luttinger liquids\cite{ilani2006} and the quantum Hall regime (in all its flavors) \cite{martin2010,venkatachalam2011}. In Ref.~\onlinecite{jarratt2020}, Jarratt et al. showed the ability to measure the van Hove singularities in a narrow GaAs quantum point contact using a resonator, demonstrating that compressibility measurements can give insight into the band structure of 1D systems. Compressibility divergence can indicate closing and reopening of the bulk gap. Compressibility measures the properties of the bulk directly, and does not rely on local probes, making it a uniquely good quantity to investigate topological phase transitions\cite{wen1990}, including the case of topological superconductors predicted to host Majorana zero modes~\cite{nozadze2016}.

\subsection{Implementation on a probe needle}
\label{probe}

As quantified in Table~\ref{tab_estimates}, sub-attofarad sensitivities can be achieved using low-Q rf resonators. In particular, LC resonators constructed from surface-mount components are commonly used for state-of-the-art charge and spin readout\cite{barthel2010,urdampilleta2019,keith2019,schupp2020}. These realizations use the self-resonance of the surface-mount inductor, and supplement it with additional capacitances to adjust the resonance frequency and characteristic impedance. While the referenced uses are demonstrated at cryogenic temperatures, these resonant circuits perform comparably well at room-temperature (Appendix~\ref{app_bonus}).

This suggests a possible realization of a rf resonator in the form of a needle probe of a micromanipulator, or a scanning probe microscope. The surrounding environment would affect both the resonance frequency and quality factor as such a probe makes contact with the device, but minimal requirements on the resonance frequency and the quality render such changes mostly irrelevant in practical use cases. Thereby the capacitance measurement method is suitable for the purpose of rapid screening of devices on a large scale.

\section{Summary}
\label{summary}

To summarize, we validate a method of measuring capacitance of micro- and nano-scale devices by means of rf resonators. The method is characterized by the sensitivity reaching values down to $75$~zF/$\sqHz$ and noise below 1~aF for moderate integration times. It is suitable for application at both room temperatures and cryogenic temperatures, including dilution refrigerators. It is suitable for measuring multiple gate capacitances with a single resonator, reducing reliance on finite element simulations for mobility measurements, can detect the quantum contribution to bulk capacitance in mesoscopic devices and can be implemented on the needle of a probe station with a micromanipulator.

\section*{Author contributions}

FKM and JVK envisioned the experiment.
LH, DdJ and JW fabricated the nanowire devices.
FKM performed the experiment and the data analysis.
DdJ and CP provided reference data on CPW resonators.
GB, SG and EPAMB grew the InSb nanowires.
YL and PK grew the InAs nanowires.
LK supervised the project.
FKM wrote the manuscipt with input from LH, DdJ, JW, CP, YL, LK and JVK.

\section*{Data availability}

Raw data, analysis code and scripts for plotting the figures for this publication are available at https://doi.org/10.5281/zenodo.5534234.

\section*{Conflicts of Interest}

FKM and JVK are named as inventors on the pending patent application, PCT Application No. PCT/EP2020/061956, which concerns the method described in this article.

\section*{Acknowledgments}

We thank John Watson and Gijs de Lange for valuable comments and suggestions.
This work was supported the Netherlands Organization for Scientific Research (NWO) and Microsoft Corporation Station Q.
FKM acknowledges support from NWO under Veni grant (VI.Veni.202.034).

\appendix
\counterwithin{figure}{section}

\section{Basic resonator model and sensitivity estimate}
\label{app_sensitivity}

The starting point for modeling the resonator is series RLC circuit coupled to a $Z_0 = 50$~$\Omega$ transmission line\cite{pozar2011}. The impedance of such a resonator is
\begin{equation}
\label{eq_raw_impedance}
	Z = R + 2 \pi i f L + \frac{1}{2 \pi i f C} \overset{\Delta f \ll \fres}{\approx} \Qe Z_0 \left( \frac{1}{\Qi} + i \frac{2 \Delta f}{\fres} \right)
\end{equation}
where $R$, $L$, and $C$ are resistance, inductance and capacitance of the RLC circuit, respectively; $\Qe = \Zc / Z_0$ is the external quality factor; $\Qi = \Zc / R$ is the internal quality factor; $Q = \left( \Qe^{-1} + \Qi^{-1} \right)^{-1}$ is the total quality factor; $\Zc = \sqrt{L/C}$ is the characteristic impedance of the resonator; $\fres = 1/(2 \pi \sqrt{LC})$ is the resonance frequency; $f$ is the probe frequency and $\Delta f = f - \fres$.

The reflection coefficient of the resonator is thereby
\begin{equation}
	\label{eq_S11}
	\S = \frac{Z-Z_0}{Z+Z_0} = 1 - \frac{ 2 Q \Qe^{-1} }{1 + 2 i Q \frac{(f - f_0) }{f_0}}.
\end{equation}

Maximum sensitivity to small changes in resonator frequency is achieved by measuring exactly on resonance
\begin{equation}
	\left. \frac{d \S}{d \fres} \right|_{f=\fres} = -4i \frac{Q^2  \Qe^{-1} }{\fres}.
\end{equation}
In a lumped-element resonator model this corresponds to a maximum sensitivity to capacitance changes of
\begin{equation}
	\left. \frac{d \S}{d C} \right|_{f=\fres} = 4 \pi i \frac{Q^2 }{\Qe} \fres \Zc.
\end{equation}
For a drive amplitude $A$ and noise voltage variance $\vn^2$ per unit of bandwidth, the sensitivity is given by
\begin{equation}
	S_C = \frac{\vn}{A} \left| \frac{d \S}{d C} \right|^{-1} .
\end{equation}

\section{Origin of the resonance asymmetry}
\label{app_asymmetric}

In the study we generally find asymmetric line shapes of the rf resonances [c.f. Fig.~\ref{fig_resonator_fit}(b)], with the asymmetry typically being more pronounced for high internal quality factors. In the hanger geometry an asymmetry is usually attributed to mismatch between impedance of input and output transmission line \cite{khalil2012}, however an such interpretation has no physical justification in a reflection measurement. One approach in reflectometry is therefore to neglect the extraction of the frequency shift and quality factor from the data. Another approach is to use additional phenomenological factors to account for asymmetry\cite{vanveen2019}. The phenomenological approach leads to correct extraction of the resonance frequency, but introduces systematic error in the extraction of the internal and external quality factors. Here, we introduce an approach utilizing a physically motivated model, that captures the resonance asymmetry.

Our model considers a cryogenic circuit depicted in Fig.~\ref{fig_resonator_fit}(a), consisting of the resonator itself, a directional coupler (characterized by a coupling parameter $\gamma$), a connecting transmission line (length $l$, and microwave propagation speed $c$) and an amplifier. We describe each of these components using scattering matrices:
\begin{align}
	S_{coupl} = &
	\begin{pmatrix}
    	0 & \sqrt{1-\gamma^2} & i\gamma & 0 \\
    	\sqrt{1-\gamma^2} & 0 & 0 & i\gamma \\
    	i\gamma & 0 & 0 & \sqrt{1-\gamma^2} \\
    	0 & i\gamma & \sqrt{1-\gamma^2} & 0
	\end{pmatrix}
	\label{eq_coupl}
	\\
	S_{transm} = &
	\begin{pmatrix}
    	0 & e^{-2 \pi i f l / c} \\
    	e^{-2 \pi i f l / c} & 0 \\
	\end{pmatrix}
	\label{eq_transm}
	\\
	S_{amp} = &
	\begin{pmatrix}
	    \sqrt{1-\alpha^2} e^{i\phi} & \alpha \\
	    \alpha & \sqrt{1-\alpha^2} e^{i\phi}
	\end{pmatrix},
	\label{eq_ampl}
\end{align}
where the amplifier is treated as a partially reflective mirror with a transmission coefficient $\alpha$, that introduces a phase shift $\phi$ to the reflected signal. The reflection coefficient of the resonator $\S$ is given by Eq.~\eqref{eq_S11}. Two effective mirrors, the resonator and the amplifier, form a low-Q cavity which modulates the transmission through the circuit from the coupled port of the directional couplet to the output of the amplifier. The modulation leads to the resonance asymmetry, and in some cases can even turn the resonance dip into peak, through the following mechanism.

The cavity formed between the resonator and the amplifier reduces the output signal, except on resonance (i.e. when on the round trip between the amplifier and the resonator the microwaves acquire a phase that is a multiple of $2 \pi$). This leads to the oscillating background in the reflection measurement. If the resonator is undercoupled ($\Qi>\Qe$), near resonance frequency $\fres$, the phase of $\S$ rapidly wraps by $2 \pi$. Therefore there must exist a frequency, close to $\fres$, for which the round trip is an exact multiple of $2 \pi$, resulting in an increase of the transmission through the cavity formed by an amplifier, and leading to an asymmetry.

Analytically, we solve a set of linear equations
\begin{align}
	\vec{V}_i^{out} = S_i \times \vec{V}_i^{in},
\end{align}
given by the scattering matrices $S_i$ (Eqs.~\eqref{eq_S11}, \eqref{eq_coupl}, \eqref{eq_transm}, \eqref{eq_ampl}), relating the microwave amplitude and phase at the inputs ($\vec{V}_i^{in}$) and outputs ($\vec{V}_i^{out}$) of each component of the circuit. In the solution we assume that the microwave drive is applied only to the coupled port of the directional coupler, and the drive is zero on the isolated port and output of the amplifier/mirror. We find the transmission from the coupled input of the directional coupler to the amplifier
\begin{align}
	\label{eq_asymmetric}
	\tilde{S}_{11} & = \frac{i \gamma \alpha \sqrt{1-\gamma^2} e^{-4 \pi f l i / c + i\phi} \S}{1 - \sqrt{1-\alpha^2} (1-\gamma^2) e^{-4 \pi f l i / c + i\phi} \S}.
\end{align}

\section{Resonator fitting}
\label{app_fit}

In a final fit to the data, we include additional prefactors to modify Eq.~\eqref{eq_asymmetric}
\begin{equation}
	\label{eq_khalil_extended}
	\tilde{\tilde{S}}_{11} = A \left( 1+B \frac{f - \fres}{\fres} \right) \times e^{-i \alpha+i \beta(f-\fres)} \times \tilde{S}_{11},
\end{equation}
and record the optimal parameters.
The term $A \left( 1+B \frac{f - \fres}{\fres} \right)$ phenomenologically accounts for a frequency-dependent attenuation and amplification, while $e^{-i \alpha+ i \beta(f-\fres)}$ accounts for the phase shift due accumulated during propagation through the transmission lines. $A$ and $B$ parametrize the background amplitude and slope, while $\alpha$ and $\beta$ parametrize the global phase shift and phase winding.

\begin{figure}[tbh]
	\includegraphics[scale=1]{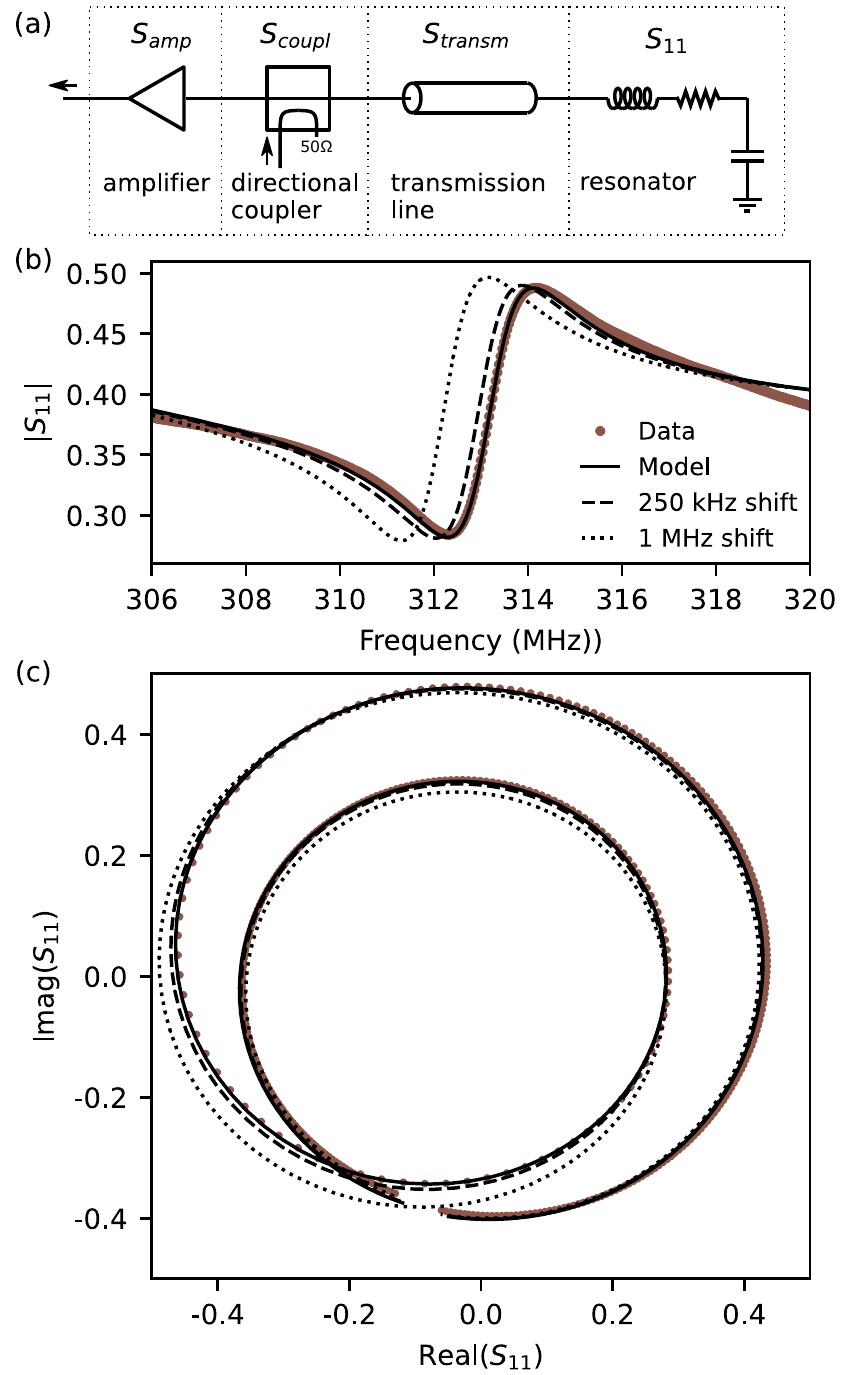}
	\caption{(a) Circuit schematics employed to model the resonance assymetry of the resonator used for characterization of the performance in Sec.~\ref{sensitivity}. (b) Absolute value of $S_{21}$ and (c) parametric plot of real and complex part. Solid line indicates the fit of Eq.~\eqref{eq_khalil_extended} to the data. Dashed and dotted line indicate the predicted signal in case the resonance frequency is lowered by 250~kHz and 1~MHz, respectively.}
	\label{fig_resonator_fit}
\end{figure}

In the fits we fix several of the parameters independently. The coupling coefficient $\gamma = 0.178$ (equivalently: 15~dB) is chosen according to the specification of the used Mini Circuits ZEDC-15-2B\footnote{https://www.minicircuits.com/WebStore/dashboard.html?model=ZEDC-15-2B}, and the reflection coefficient $\alpha = 0.398$ corresponds to 8~dB return loss of the Cosmic Microwave CITLF2 HEMT cryogenic amplifier\footnote{https://www.cosmicmicrowavetechnology.com/citlf2}. We set the value of $(2 l / c)^{-1} = 111.4$~MHz based on the measurement of the reflection at 4K, with no device mounted in the setup. The remaining parameters are optimized in the nonlinear fit. Fig.~\ref{fig_resonator_fit} depicts the fit result for the resonator used to quantify sensitivity in Sec.~\ref{sensitivity}.

\section{Frequency shift from fixed-frequency measurement}
\label{app_fixed_freq}

To maximize the measurement sensitivity it is optimal to perform a fixed-frequency measurement, near the resonance frequency, for which the reflection coefficient responds most strongly. To recover the frequency shift from such a fixed-frequency measurement we perform a calibration resonator measurement (Fig.~\ref{fig_resonator_fit}) and fit the analytical model (Eq.~\eqref{eq_khalil_extended}) to the data (Appendix.~\ref{app_fit}). We fix all of the parameters of the model, except for the resonance frequency $\fres$. In this way we are able to predict the expected reflection coefficient for different values of resonance frequency (e.g. dashed and dotted lines in Fig.~\ref{fig_resonator_fit}).

Next, we measure the reflection at fixed frequency $f$ versus the gate voltage. For each data point we perform numerical optimization to find $f_0$ which best matches the data, and identify the corresponding value as the resonance frequency at a given gate voltage.

In this work we assume that the internal quality factor $\Qi$ of the resonator is not gate-dependent, which is reasonably fulfilled [Fig.~\ref{fig_comparison}(b)]. We note that since the reflection coefficient is complex-valued it could be used to infer two real-valued parameters simultaneously ($\fres$ and $\Qi$) via numerical optimization.

\section{Additional data sets}
\label{app_bonus}

In this appendix we present additional data sets, backing up the numerical values provided in the main text.

\begin{figure*}
	\includegraphics[scale=1]{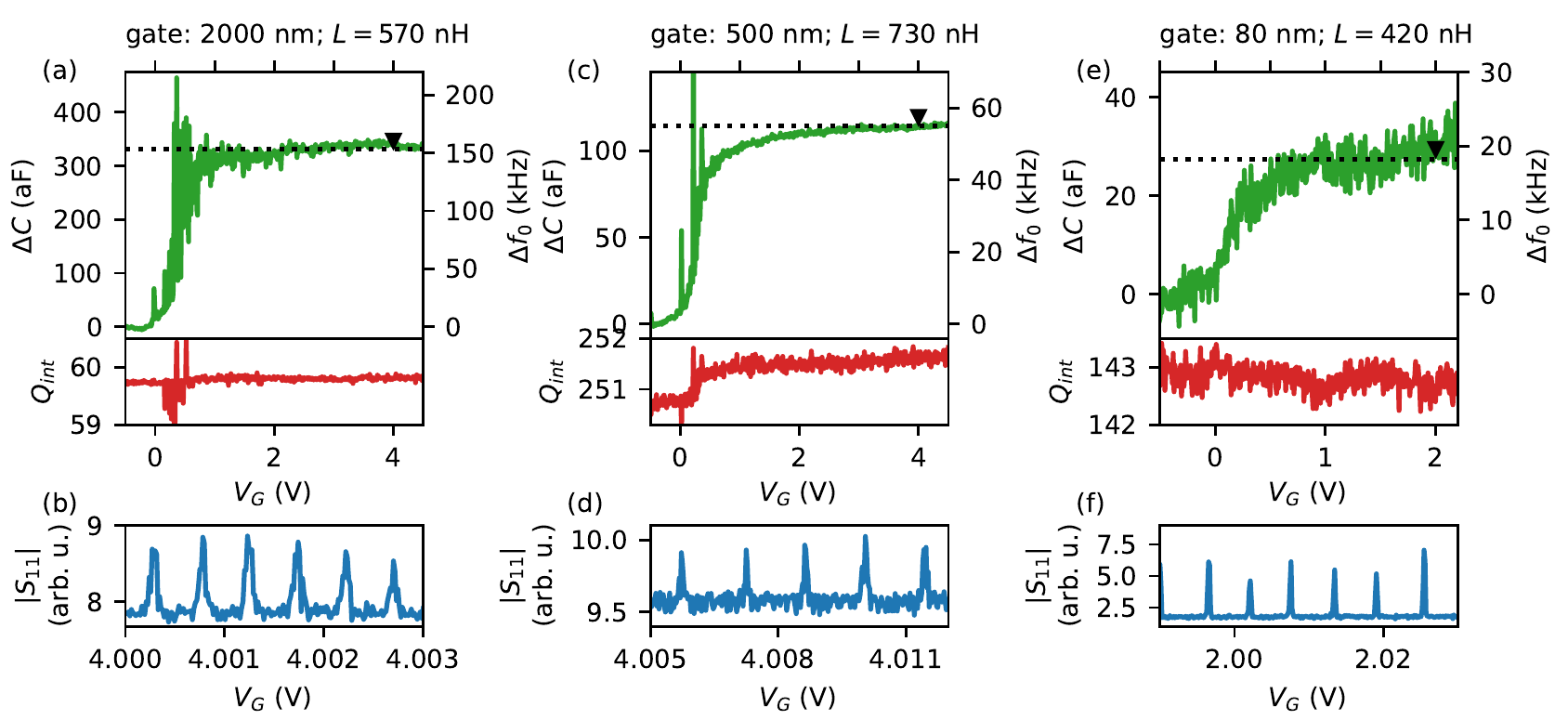}
	\caption{(a,c,e) C-V measurements using the rf resonators for the three InSb devices with the gate width of 2000, 500 and 80 nm. Bottom panels present internal quality factors of the resonators as a function of the gate voltage $V_G$. (b,d,f) Rf-conductance measurements of the quantum dots formed in the nanowire devices.}
	\label{fig_three_capacitances}
\end{figure*}

Figure~\ref{fig_three_capacitances} presents the CV measurement for three devices listed in Table~\ref{tab_comparison}. In the top panel of Fig.~\ref{fig_three_capacitances} the dashed line indicates the capacitance value extracted from the periodicity of Coulomb blockade, and the triangular marker indicates the gate voltage $V_G$ which was used for tuning the quantum dot.

\begin{figure}
	\includegraphics[scale=1]{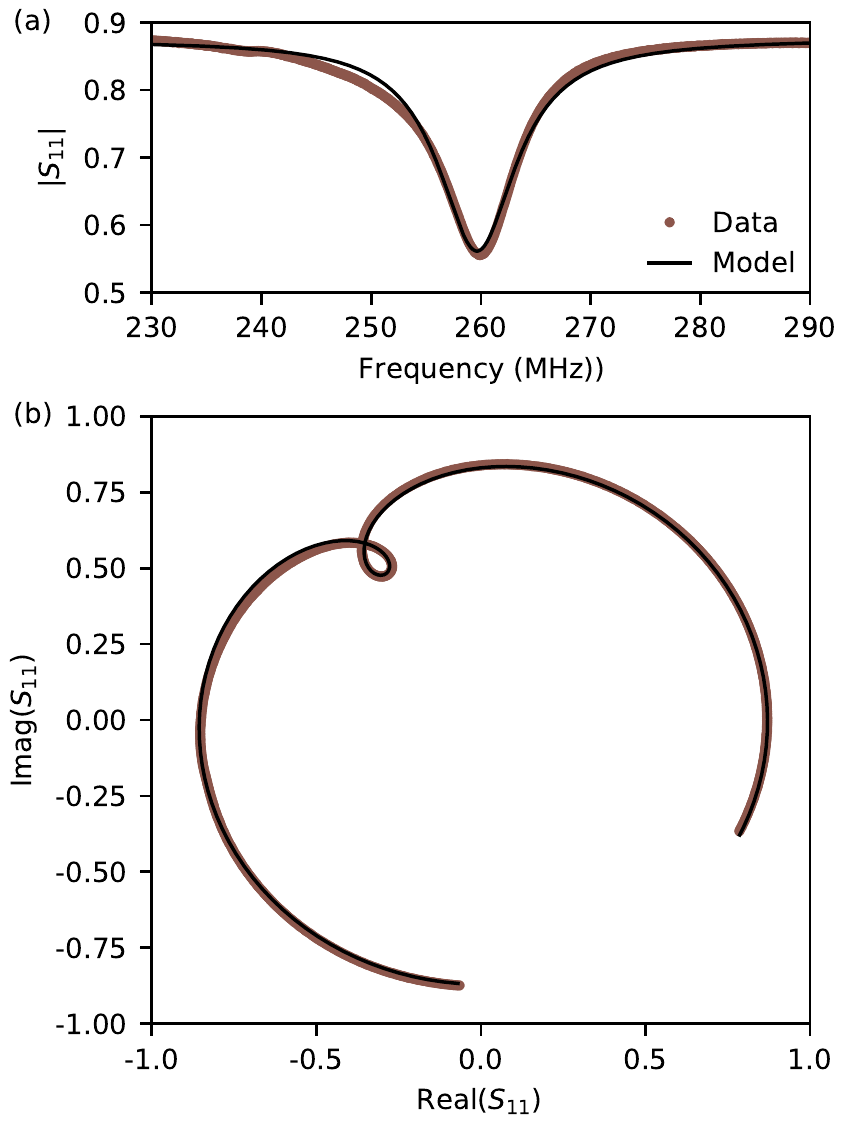}
	\caption{(a) Magnitude of the reflection around the self-resonance frequency of the surface mount inductor. (b) Parametric plot of the real and imaginary part of the reflection around the self resonance-frequency. Black lines indicate the fit to the data.}
	\label{fig_RT}
\end{figure}

Figure~\ref{fig_RT} presents the measurement of the self-resonance of the 1~$\mu$H surface-mount inductor (Coilcraft, 1008CS-102X\_E\_), together with a complex fit to the data. In these measurements no directional coupler or cryo-amplifier was used, therefore the data is fited by Eq.~\ref{eq_S11}, with the prefactors listed in Appendix~\ref{app_fit}. The parameters extracted from the fit were used in the final entry in Table~\ref{tab_estimates}.

\begin{figure}
	\includegraphics[scale=1]{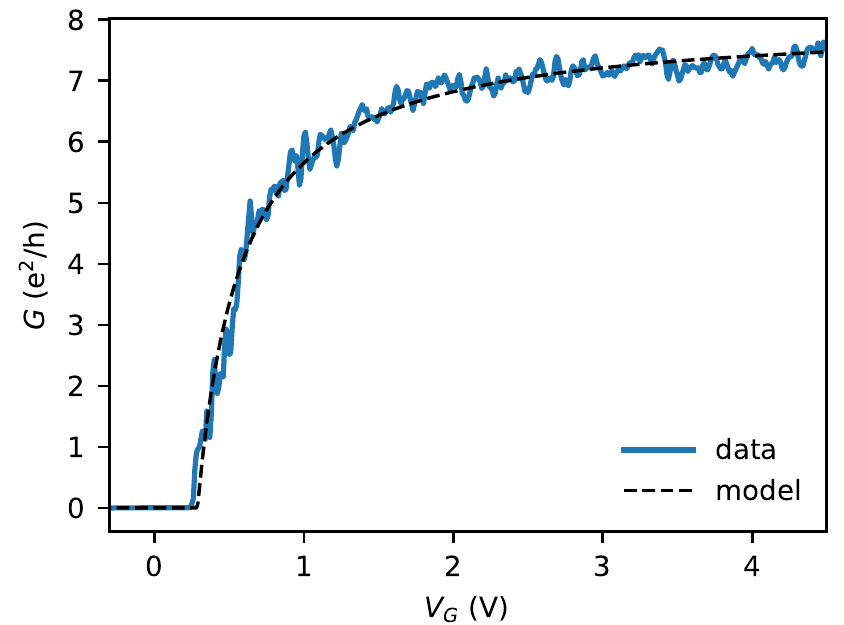}
	\caption{Pinch off curve of the 500 nm InSb device, used for extraction of mobility in Fig.~\ref{fig_mobility}.}
	\label{fig_field-effect}
\end{figure}

Figure~\ref{fig_field-effect} shows a fit of the field-effect model~\cite{gul2015} to the pinch-off curve from Fig.~\ref{fig_mobility}. This fit yields the quoted value of mobility $\mu_{FE} = 2.3 \times 10^4$~cm$^2$/Vs.

\bibliography{biblio}

\end{document}